\documentclass[12pt]{article}
\usepackage{graphicx}

\def\hybrid{\topmargin 0pt      \oddsidemargin 0pt
        \headheight 0pt \headsep 0pt
       \voffset-1cm
        \textwidth 6.25in       
       \textheight 9.5in       
        \marginparwidth 0.0in
        \parskip 5pt plus 1pt   \jot = 1.5ex}
\catcode`\@=11
\def\marginnote#1{}

\newcount\hour
\newcount\minute
\newtoks\amorpm
\hour=\time\divide\hour by60
\minute=\time{\multiply\hour by60 \global\advance\minute by-\hour}
\edef\standardtime{{\ifnum\hour<12 \global\amorpm={am}%
        \else\global\amorpm={pm}\advance\hour by-12 \fi
        \ifnum\hour=0 \hour=12 \fi
        \number\hour:\ifnum\minute<10 0\fi\number\minute\the\amorpm}}
\edef\militarytime{\number\hour:\ifnum\minute<10 0\fi\number\minute}

\def\draftlabel#1{{\@bsphack\if@filesw {\let\thepage\relax
   \xdef\@gtempa{\write\@auxout{\string
      \newlabel{#1}{{\@currentlabel}{\thepage}}}}}\@gtempa
   \if@nobreak \ifvmode\nobreak\fi\fi\fi\@esphack}
        \gdef\@eqnlabel{#1}}
\def\@eqnlabel{}
\def\@vacuum{}
\def\draftmarginnote#1{\marginpar{\raggedright\scriptsize\tt#1}}

\def\draftlabel#1{{\@bsphack\if@filesw {\let\thepage\relax
   \xdef\@gtempa{\write\@auxout{\string
      \newlabel{#1}{{\@currentlabel}{\thepage}}}}}\@gtempa
   \if@nobreak \ifvmode\nobreak\fi\fi\fi\@esphack}
        \gdef\@eqnlabel{#1}}
\def\@eqnlabel{}
\def\@vacuum{}
\def\draftmarginnote#1{\marginpar{\raggedright\scriptsize\tt#1}}

\def\draft{\oddsidemargin -.5truein
        \def\@oddfoot{\sl preliminary draft \hfil
        \rm\thepage\hfil\sl\today\quad\militarytime}
        \let\@evenfoot\@oddfoot \overfullrule 3pt
        \let\label=\draftlabel
        \let\marginnote=\draftmarginnote
   \def\@eqnnum{(\theequation)\rlap{\kern\marginparsep\tt\@eqnlabel}%
\global\let\@eqnlabel\@vacuum}  }

\def\numberbysection{\@addtoreset{equation}{section}
        \def\theequation{\thesection.\arabic{equation}}}

\def\underline#1{\relax\ifmmode\@@underline#1\else
        $\@@underline{\hbox{#1}}$\relax\fi}

\def\titlepage{\@restonecolfalse\if@twocolumn\@restonecoltrue\onecolumn
     \else \newpage \fi \thispagestyle{empty}\c@page\z@
        \def\thefootnote{\fnsymbol{footnote}} }

\def\endtitlepage{\if@restonecol\twocolumn \else  \fi
        \def\thefootnote{\arabic{footnote}}
        \setcounter{footnote}{0}}  
\relax

\numberbysection
\hybrid

\newfont{\Bbb}{msbm10 scaled 1\@ptsize00}
\newfont{\Bbbb}{msbm7 scaled 1\@ptsize00}

\newcommand{\DDD}{\raise-1pt\hbox{$\mbox{\Bbbb D}$}}

\newcommand{\UUU}{\raise-1pt\hbox{$\mbox{\Bbbb U}$}}

\newcommand{\ZZ}{\mbox{\Bbb Z}}
\newcommand{\z}{\raise-1pt\hbox{$\mbox{\Bbbb Z}$}}

\newcommand{\sss}{\raise-1pt\hbox{$\mbox{\Bbbb S}$}}

\def\beq{\begin{equation}}
\def\eeq{\end{equation}}
\def\p{\partial}

\newtheorem{lemma-definition}{Lemma-Definition}[section]

\begin{document}

\begin{titlepage}

\title{Dispersionless version of the multicomponent KP hierarchy
revisited}

\author{A. Zabrodin\thanks{Skolkovo Institute of Science and
Technology, 143026, Moscow, Russia and
National Research University Higher School of
Economics,
20 Myasnitskaya Ulitsa, Moscow 101000, Russia, 
and NRC ``Kurchatov institute'', Moscow, Russia;
e-mail: zabrodin@itep.ru}}

\date{April 2024}

\maketitle

\vspace{-7cm} \centerline{ \hfill ITEP-TH-11/24}\vspace{7cm}

\begin{abstract}

We revisit dispersionless version of the multicomponent KP hierarchy
considered previously by Takasaki and Takebe. In contrast to their
study, we do not fix any distinguished component treating all of them
on equal footing. We obtain nonlinear equations for dispersionless 
tau-function (the $F$-function) and represent them using the trigonometric
parametrization. In this trigonometric uniformization the equations considerably simplify and acquire a nice form.

\end{abstract}

\end{titlepage}

\vspace{5mm}

%

\tableofcontents

\vspace{5mm}

\section{Introduction}

The multicomponent KP hierarchy was introduced in the paper \cite{DJKM81}
in 1981.
Since then it was considered in the works \cite{KL93,TT07,Teo11}. 
Recently, some generalization of it was suggested in \cite{KZ23}
under the name of the universal hierarchy. (The difference of the
approach in \cite{KZ23} with the
previous works is that the integer variables of the hierarchy are
allowed to take arbitrary complex values, with the equations in these
variables being still difference.) 

In this note we are mainly interested in the dispersionless limit
of the hierarchy, tending the small ``dispersion parameter'' $\hbar$
to zero. The general approach to dispersionless limits of
integrable hierarchies was developed in \cite{TT95} by Takasaki and Takebe.
In \cite{TT07}, they analyzed the dispersionless limit of the 
multicomponent KP hierarchy. In their approach, one of the components
was distinguished. We find it appropriate to revisit 
this problem. In our approach, all components are treated on equal
footing. We obtain nonlinear equations for the dispersionless limit
of (logarithm of) tau-function (the $F$-function) 
which follow from the bilinear
equations of the Hirota-Miwa type. 

The main result of this paper is a new form of the dispersionless
equations for the $F$-function which emerges after a trigonometric
parametrization. In this form, the equations considerably simplify
and some of them become equivalent, so the number of independent
equations reduces (basically, only one main equation is left).

The paper is organized as follows. In section 2 we present the 
generating bilinear integral equation for the tau-function and derive
different equations of the Hirota-Miwa type as corollaries of it. 
Section 3 is devoted to the dispersionless limit. We obtain the 
dispersionless version of all the Hirota-Miwa equations from the previous
section. In section 4 we suggest the trigonometric parametrization and 
rewrite all the equations obtained in the previous section in the nice
trigonometric form. Section 5 contains concluding remarks. 

\section{Bilinear equations for the tau-function}

Let us start from the $n$-component KP hierarchy 
extended by certain additional integrable flows
\cite{DJKM81,KL93,TT07,Teo11}. In \cite{KZ23} it is called the 
universal hierarchy.
The independent variables are $n$ infinite sets of (in general complex) 
``times''
$$
{\bf t}=\{{\bf t}_1, {\bf t}_2, \ldots , {\bf t}_n\}, \qquad
{\bf t}_{\alpha}=\{t_{\alpha , 1}, t_{\alpha , 2}, 
t_{\alpha , 3}, \ldots \, \},
\qquad \alpha = 1, \ldots , n
$$
and $n$ additional variables $s_1, \ldots , s_n$
such that 
\beq\label{s1}
\sum_{\alpha =1}^n s_{\alpha}=0.
\eeq
By ${\bf s}$ we denote the vector
${\bf s}=\{s_1, \ldots , s_n\}$ and by ${\bf e}_{\alpha}$ the vector
whose $\alpha$'s component is 1 and all other components 0.
We will also use the following standard
notation:
\beq\label{st1}
\left ({\bf t}\pm [z^{-1}]_{\gamma}\right )_{\alpha j}=t_{\alpha , j}\pm
\delta_{\alpha \gamma} \frac{z^{-j}}{j},
\eeq
\beq\label{st2}
\xi ({\bf t}_{\alpha}, z)=\sum_{j\geq 1}t_{\alpha , j}z^j.
\eeq
In general we treat $s_1, \ldots , s_n$ as complex variables,
as in \cite{KZ23}.
If they are restricted to be integers, the hierarchy coincides with the 
one considered in
\cite{DJKM81}--\cite{Teo11}.

In the bilinear formalism, 
the dependent variable is the tau-function 
$\tau ({\bf s}, {\bf t})$ 
The universal hierarchy is the infinite set of bilinear equations
for the tau-function which are encoded in the basic bilinear
relation \cite{DJKM81,Teo11}
\beq\label{s3}
\begin{array}{l}
\displaystyle{
\sum_{\gamma =1}^n \epsilon_{\alpha \gamma}({\bf s})
\epsilon^{-1}_{\beta \gamma}({\bf s}')
\oint_{C_{\infty}}\! dz \, 
z^{s_{\gamma}-s_{\gamma}'+\delta_{\alpha \gamma}+\delta_{\beta \gamma}-2}
e^{\xi ({\bf t}_{\gamma}-{\bf t}_{\gamma}', \, z)}}
\\ \\
\displaystyle{\phantom{aaaaaaaaaa}
\times \tau \left ({\bf s}+{\bf e}_{\alpha}-{\bf e}_{\gamma}, 
{\bf t}-[z^{-1}]_{\gamma}\right )
\tau \left ({\bf s}'+{\bf e}_{\gamma}-{\bf e}_{\beta}, {\bf t}'+[z^{-1}]_{\gamma}\right )=0}
\end{array}
\eeq 
valid for any ${\bf t}$, ${\bf t}'$, ${\bf s}$, ${\bf s}'$
such that ${\bf s}-{\bf s}' \in \ZZ ^n$
(and subject to the constraint (\ref{s1})). 
In (\ref{s3}) 
\beq\label{s3a}
\epsilon_{\alpha \gamma}({\bf s})=\left \{
\begin{array}{cl} 
\displaystyle{\exp \, \Bigl (-i\pi \!\! \sum_{\alpha <\mu \leq \gamma}
\!\! s_{\mu}
\Bigr )}, 
&\quad \alpha < \gamma
\\ 
\hspace{-1.5cm}1, &\quad \alpha =\gamma
\\ 
\displaystyle{-\, \vphantom{\sum^{\alpha \leq }}
\exp \, \Bigl (i\pi \!\! \sum_{\gamma <\mu \leq \alpha}\!\! s_{\mu}
\Bigr )}, 
&\quad \alpha > \gamma 
\end{array}
\right.
\eeq
(see \cite{KZ23}).
The contour $C_{\infty}$ is a big circle around infinity.
It is easy to see that for ${\bf s}-{\bf s}'\in \ZZ^n$ the factor
$\epsilon_{\alpha \gamma}({\bf s})
\epsilon^{-1}_{\beta \gamma}({\bf s}')$ multiplied by 
$\epsilon_{\alpha \beta}({\bf s})$
is just a sign factor $\pm 1$ depending only on ${\bf s}-{\bf s}'$.

Different bilinear relations 
for the tau-function of the Hirota-Miwa type which follow from
(\ref{s3}) for special choices of ${\bf s}-{\bf s}'$ and ${\bf t}-{\bf t}'$
are given in \cite{Teo11}. We present some of them below, with a sketch
of the derivation.

First of all we consider (\ref{s3}) at $\alpha =\beta$, differentiate it
with respect to $t_{\alpha , 1}$ and put ${\bf s}'={\bf s}$,
${\bf t}-{\bf t}'=[a^{-1}]_{\alpha}+[b^{-1}]_{\alpha}$, so that
$$
e^{\xi ({\bf t}_{\alpha}-{\bf t}'_{\alpha}, z)}=
\frac{a\, b}{(a-z)(b-z)}, \quad
e^{\xi ({\bf t}_{\gamma}-{\bf t}'_{\gamma}, z)}=1 \quad
\mbox{for $\gamma \neq \alpha$}.
$$ 
We get
$$
\oint_{C_{\infty}}dz \frac{abz}{(a-z)(b-z)}\, \tau ({\bf s},
{\bf t}-[z^{-1}]_{\alpha})\tau ({\bf s},
{\bf t}-[a^{-1}]_{\alpha}-[b^{-1}]_{\alpha}+[z^{-1}]_{\alpha})
$$
$$
+\oint_{C_{\infty}}dz \frac{ab}{(a-z)(b-z)}\, \p_{t_{\alpha ,1}}
\tau ({\bf s},
{\bf t}-[z^{-1}]_{\alpha})\tau ({\bf s},
{\bf t}-[a^{-1}]_{\alpha}-[b^{-1}]_{\alpha}+[z^{-1}]_{\alpha})=0.
$$
The residue calculus followed by some simple transformations yields
the equation
\beq\label{s4}
\frac{\tau ({\bf s}, {\bf t})\tau ({\bf s}, {\bf t}+
[a^{-1}]_{\alpha}+[b^{-1}]_{\alpha})}{\tau ({\bf s}, {\bf t}+
[a^{-1}]_{\alpha})\tau ({\bf s}, {\bf t}+
[b^{-1}]_{\alpha})}=
1-\frac{1}{a-b}\, \p_{t_{\alpha ,1}}\log \frac{\tau ({\bf s}, {\bf t}+
[a^{-1}]_{\alpha})}{\tau ({\bf s}, {\bf t}+
[b^{-1}]_{\alpha})}.
\eeq
Now consider (\ref{s3}) with $\alpha =\beta$ and ${\bf s}'={\bf s}
+{\bf e}_{\alpha}-{\bf e}_{\mu}$ (with $\mu \neq \alpha$) and
${\bf t}-{\bf t}'=[a^{-1}]_{\alpha}+[b^{-1}]_{\alpha}$. The residue
calculus yields:
\beq\label{s5}
\begin{array}{c}
\displaystyle{
\frac{\tau ({\bf s}, {\bf t})\tau ({\bf s}
+{\bf e}_{\alpha}-{\bf e}_{\beta}, {\bf t}+
[a^{-1}]_{\alpha}+[b^{-1}]_{\alpha})}{\tau ({\bf s}, {\bf t}+
[a^{-1}]_{\alpha})\tau ({\bf s}+
{\bf e}_{\alpha}-{\bf e}_{\beta}, {\bf t}+
[b^{-1}]_{\alpha})}}
\\ \\
\displaystyle{
=\frac{1}{a-b}\left (a\frac{\tau ({\bf s}
+{\bf e}_{\alpha}-{\bf e}_{\beta}, 
{\bf t}+[b^{-1}]_{\alpha})}{\tau ({\bf s}, {\bf t}+
[b^{-1}]_{\alpha})}-b\frac{\tau ({\bf s}
+{\bf e}_{\alpha}-{\bf e}_{\beta}, 
{\bf t}+[a^{-1}]_{\alpha})}{\tau ({\bf s}, {\bf t}+
[a^{-1}]_{\alpha})}\right )},
\end{array}
\eeq
where we have changed $\mu \to \beta$. 
The next choice in (\ref{s3}) with $\alpha =\beta$ is ${\bf s}'={\bf s}$
and ${\bf t}-{\bf t}'=[a^{-1}]_{\alpha}+[b^{-1}]_{\mu}$ 
($\mu \neq \alpha$). The residue calculus yields the equation
\beq\label{s6}
\begin{array}{c}
\tau ({\bf s}, {\bf t})\tau ({\bf s}, {\bf t}+
[a^{-1}]_{\alpha}+[b^{-1}]_{\beta})-
\tau ({\bf s}, {\bf t}+[a^{-1}]_{\alpha})\tau ({\bf s}, {\bf t}+
[b^{-1}]_{\beta})
\\ \\
=(ab)^{-1}\tau ({\bf s}
+{\bf e}_{\alpha}-{\bf e}_{\beta}, {\bf t}+[a^{-1}]_{\alpha})
\tau ({\bf s}
+{\bf e}_{\beta}-{\bf e}_{\alpha}, {\bf t}+[b^{-1}]_{\beta}), 
\end{array}
\eeq
where we again have changed $\mu \to \beta$. 

Let us now consider the case $\alpha \neq \beta$ in (\ref{s3}). First,
we apply $\p_{t_{\alpha ,1}}$ and put ${\bf s}'={\bf s}$,
${\bf t}-{\bf t}'=[a^{-1}]_{\alpha}+[b^{-1}]_{\beta}$ after that.
The residue calculus yields:
\beq\label{s7}
a\frac{\tau ({\bf s}, {\bf t}+
[a^{-1}]_{\alpha}+
[b^{-1}]_{\beta}))\tau ({\bf s}
+{\bf e}_{\alpha}-{\bf e}_{\beta}, {\bf t})}{\tau ({\bf s}, {\bf t}+
[b^{-1}]_{\beta}))
\tau ({\bf s}
+{\bf e}_{\alpha}-{\bf e}_{\beta}, {\bf t}+[a^{-1}]_{\alpha})}=a+
\p_{t_{\alpha ,1}}\log \frac{\tau ({\bf s}, {\bf t}+
[b^{-1}]_{\beta})}{\tau ({\bf s}
+{\bf e}_{\alpha}-{\bf e}_{\beta}, {\bf t}+[a^{-1}]_{\alpha})}.
\eeq
The next choice is ${\bf s}'={\bf s}$, 
${\bf t}-{\bf t}'=[a^{-1}]_{\mu}$ with $\mu \neq \alpha , \beta$.
In this case equation (\ref{s3}) yields:
\beq\label{s8}
\begin{array}{c}
\tau ({\bf s}, {\bf t})
\tau ({\bf s}+{\bf e}_{\alpha}-{\bf e}_{\beta}, {\bf t}+[a^{-1}]_{\mu})-
\tau ({\bf s}, {\bf t}+[a^{-1}]_{\mu})\tau ({\bf s}
+{\bf e}_{\alpha}-{\bf e}_{\beta}, {\bf t})
\\ \\
+\, \epsilon_{\alpha \beta}\epsilon_{\alpha \mu}\epsilon_{\beta \mu}
a^{-1}\tau ({\bf s}
+{\bf e}_{\alpha}-{\bf e}_{\mu}, {\bf t})
\tau ({\bf s}
+{\bf e}_{\mu}-{\bf e}_{\beta}, {\bf t}+[a^{-1}]_{\mu})=0,
\end{array}
\eeq
where $\epsilon_{\alpha \beta}=1$ at $\alpha \leq \beta$ and 
$\epsilon_{\alpha \beta}=-1$ at $\alpha > \beta$.
Finally, we put ${\bf s}'={\bf s}+{\bf e}_{\alpha}-{\bf e}_{\beta}$,
${\bf t}-{\bf t}'=[a^{-1}]_{\alpha}+[b^{-1}]_{\alpha}$
in (\ref{s3}). The residue calculus yields:
\beq\label{s9}
\begin{array}{l}
\displaystyle{
\frac{a}{b}\, \frac{\tau ({\bf s}-
{\bf e}_{\alpha}+{\bf e}_{\beta}, {\bf t}+[a^{-1}]_{\alpha})
\tau ({\bf s}+
{\bf e}_{\alpha}-{\bf e}_{\beta}, {\bf t}+[b^{-1}]_{\alpha})}{\tau
({\bf s}, {\bf t})\tau ({\bf s}, {\bf t})+[a^{-1}]_{\alpha}+
[b^{-1}]_{\alpha})}}
\\ \\
\displaystyle{\phantom{aaaaaaaaaaaaa}-
\frac{b}{a}\, \frac{\tau ({\bf s}-
{\bf e}_{\alpha}+{\bf e}_{\beta}, {\bf t}+[b^{-1}]_{\alpha})
\tau ({\bf s}+
{\bf e}_{\alpha}-{\bf e}_{\beta}, {\bf t}+[a^{-1}]_{\alpha})}{\tau
({\bf s}, {\bf t})\tau ({\bf s}, {\bf t})+[a^{-1}]_{\alpha}+
[b^{-1}]_{\alpha})}}
\\ \\
\displaystyle{ =(a-b)\p_{t_{\beta ,1}}\log \frac{\tau ({\bf s}, {\bf t})+[a^{-1}]_{\alpha}+
[b^{-1}]_{\alpha})}{\tau
({\bf s}, {\bf t})}}.
\end{array}
\eeq

Another equation can be obtained in the following way.
Put $\alpha =\beta =\mu$ in (\ref{s3}):
$$
\begin{array}{l}
\displaystyle{
\sum_{\gamma =1}^n \epsilon_{\mu \gamma}({\bf s})
\epsilon^{-1}_{\mu \gamma}({\bf s}')
\oint_{C_{\infty}}\! dz \, 
z^{s_{\gamma}-s_{\gamma}'+2\delta_{\mu \gamma}-2}
e^{\xi ({\bf t}_{\gamma}-{\bf t}_{\gamma}', \, z)}}
\\ \\
\displaystyle{\phantom{aaaaaaaaaa}
\times \tau \left ({\bf s}+{\bf e}_{\mu}-{\bf e}_{\gamma}, 
{\bf t}-[z^{-1}]_{\gamma}\right )
\tau \left ({\bf s}'+{\bf e}_{\gamma}-{\bf e}_{\mu}, {\bf t}'+[z^{-1}]_{\gamma}\right )=0},
\end{array}
$$ 
and put here ${\bf s}-{\bf s}'={\bf e}_{\alpha}+{\bf e}_{\beta}-
2{\bf e}_{\mu}$, ${\bf t}-{\bf t}'=[a^{-1}]_{\alpha}+[b^{-1}]_{\beta}$
with $\alpha \neq \beta$ and $\mu \neq \alpha , \beta$. 
The residue calculus yields:
$$
\epsilon_{\mu \alpha}({\bf s})\epsilon_{\mu \alpha}^{-1}
({\bf s}+2{\bf e}_{\mu}-{\bf e}_{\alpha}-{\bf e}_{\beta})
\p_{t_{\mu ,1}}\tau ({\bf s}+{\bf e}_{\mu}-{\bf e}_{\alpha}, 
{\bf t}+[b^{-1}]_{\beta})
\tau ({\bf s}+{\bf e}_{\mu}-{\bf e}_{\beta}, 
{\bf t}+[a^{-1}]_{\alpha})
$$
$$
+\epsilon_{\mu \beta}({\bf s})\epsilon_{\mu \beta}^{-1}
({\bf s}+2{\bf e}_{\mu}-{\bf e}_{\alpha}-{\bf e}_{\beta})
\p_{t_{\mu ,1}}\tau ({\bf s}+{\bf e}_{\mu}-{\bf e}_{\beta}, 
{\bf t}+[a^{-1}]_{\alpha})
\tau ({\bf s}+{\bf e}_{\mu}-{\bf e}_{\alpha}, 
{\bf t}+[b^{-1}]_{\beta})
$$
$$
+\tau ({\bf s}, {\bf t}+[a^{-1}]_{\alpha}+[b^{-1}]_{\beta})
\tau ({\bf s}+2{\bf e}_{\mu}-{\bf e}_{\alpha}-{\bf e}_{\beta},
{\bf t})=0.
$$
The case-by-case inspection shows that
$$
\epsilon_{\mu \alpha}({\bf s})\epsilon_{\mu \alpha}^{-1}
({\bf s}+2{\bf e}_{\mu}-{\bf e}_{\alpha}-{\bf e}_{\beta})=-
\epsilon_{\mu \beta}({\bf s})\epsilon_{\mu \beta}^{-1}
({\bf s}+2{\bf e}_{\mu}-{\bf e}_{\alpha}-{\bf e}_{\beta})=
-\epsilon_{\alpha \beta}\epsilon_{\alpha \mu}\epsilon_{\beta \mu}.
$$
Therefore, the equation finally reads:
\beq\label{s10}
\begin{array}{l}
\displaystyle{
\epsilon_{\alpha \beta}
\frac{\tau ({\bf s}+{\bf e}_{\beta}-{\bf e}_{\mu}, 
{\bf t}+[a^{-1}]_{\alpha}+[b^{-1}]_{\beta})
\tau ({\bf s}+{\bf e}_{\mu}-{\bf e}_{\alpha}, 
{\bf t})}{\tau ({\bf s}+{\bf e}_{\beta}-{\bf e}_{\alpha}, 
{\bf t}+[b^{-1}]_{\beta})\tau ({\bf s}, {\bf t}+[a^{-1}]_{\alpha})}}
\\ \\
\displaystyle{ \phantom{aaaaaaaaaaaaaaaaaaaa}
=\epsilon_{\alpha \mu}\epsilon_{\beta \mu}
\p_{t_{\mu ,1}}\log
\frac{\tau ({\bf s}+{\bf e}_{\beta}-{\bf e}_{\alpha}, 
{\bf t}+[b^{-1}]_{\beta})}{\tau ({\bf s}, {\bf t}+[a^{-1}]_{\alpha})}.}
\end{array}
\eeq

\section{The dispersionless limit}

In order to perform the dispersionless limit \cite{TT95}, 
one should introduce a small
parameter $\hbar$ and rescale the times ${\bf t}$ 
and variables ${\bf s}$ as
$t_{\alpha , k}\to t_{\alpha , k}/\hbar$, 
$s_{\alpha}\to s_{\alpha}/\hbar$. Introduce a function 
$F({\bf s}, {\bf t}; \hbar )$ related to the tau-function by the
formula
\beq\label{d1}
\tau ({\bf s}/\hbar , {\bf t}/\hbar )=\exp \Bigl ( \frac{1}{\hbar^2}\,
F({\bf s}, {\bf t}; \hbar )\Bigr )
\eeq
and consider the limit $F=\lim\limits_{\hbar \to 0}
F({\bf s}, {\bf t}; \hbar )$. The function $F$ represents the 
tau-function in the dispersionless limit $\hbar \to 0$. It satisfies 
an infinite number of nonlinear differential equations which follow
from the bilinear equations for the tau-function. In the
dispersionless limit the difference equations in
the variables $s_{\alpha}$ become differential, and the
derivative $\p_{s_{\alpha}}$ will be denoted simply as $\p_{\alpha}$.
Introduce also the differential operators 
\beq\label{d2}
D_{\alpha}(z)=\sum_{k\geq 1} \frac{z^{-k}}{k}\, \p_{t_{\alpha , k}},
\eeq
so that
$$
\tau ({\bf s}/\hbar , {\bf t}/\hbar +[z^{-1}]_{\alpha})
=\exp \Bigl ( \frac{1}{\hbar^2}\,
e^{\hbar D_{\alpha}(z)}
F({\bf s}, {\bf t}; \hbar )\Bigr )
$$
and
$$
\tau ({\bf s}/\hbar +{\bf e}_{\alpha}-{\bf e}_{\beta},
{\bf t}/\hbar )=\exp \Bigl ( \frac{1}{\hbar^2}\,
e^{\hbar \p_{\alpha}-
\hbar \p_{\beta}}F({\bf s}, {\bf t}; \hbar )\Bigr ).
$$

Let us obtain dispersionless versions of equations (\ref{s4})--(\ref{s9}).
Equation (\ref{s4}) can be rewritten as
$$
\exp \left (\frac{1}{\hbar^2}\Bigl ( 1+e^{\hbar 
D_{\alpha}(a)+\hbar D_{\alpha}(b)}-e^{\hbar 
D_{\alpha}(a)}-e^{\hbar D_{\alpha}(b)}\Bigr )F\right )=1-
\frac{\hbar^{-1}}{a-b}\, \p_{t_{\alpha ,1}}\Bigl (
e^{\hbar D_{\alpha}(a)}-e^{\hbar D_{\alpha}(b)}\Bigr )F.
$$
In this form it is ready for the dispersionless limit $\hbar \to 0$
which yields:
\beq\label{s4a}
e^{D_{\alpha}(a)D_{\alpha}(b)F}=1-\frac{D_{\alpha}(a)\p_{t_{\alpha ,1}}F
-D_{\alpha}(b)\p_{t_{\alpha ,1}}F}{a-b}.
\eeq
The limits of the other equations can be found in a similar way. 
They are as follows. The limit of equation (\ref{s5}):
\beq\label{s5a}
e^{D_{\alpha}(a)D_{\alpha}(b)F}=\frac{a
e^{-D_{\alpha}(a)(\p_{\alpha}-\p_{\beta})F}-b
e^{-D_{\alpha}(b)(\p_{\alpha}-\p_{\beta})F}}{a-b}.
\eeq
The limit of equation (\ref{s6}):
\beq\label{s6a}
e^{D_{\alpha}(a)D_{\beta}(b)F}=1+\frac{1}{ab}\,
e^{(D_{\alpha}(a)-D_{\beta}(b))(\p_{\alpha}-\p_{\beta})F +
(\p_{\alpha}-\p_{\beta})^2F}.
\eeq
The limit of equation (\ref{s7}):
\beq\label{s7a}
ae^{D_{\alpha}(a)D_{\beta}(b)F-D_{\alpha}(a)(\p_{\alpha}-\p_{\beta})F}=
a-(\p_{\alpha}-\p_{\beta}+D_{\alpha}(a)-D_{\beta}(b))\p_{t_{\alpha ,1}}F.
\eeq
The limit of equation (\ref{s8}):
\beq\label{s8a}
e^{D_{\mu}(a)(\p_{\alpha}-\p_{\beta})F}-1+
\epsilon_{\alpha \beta}\epsilon_{\alpha \mu}\epsilon_{\beta \mu}
a^{-1}e^{D_{\mu}(a)(\p_{\mu}-\p_{\beta})F +(\p_{\mu}-\p_{\alpha})
(\p_{\mu}-\p_{\beta})F}=0.
\eeq
The limit of equation (\ref{s9}):
\beq\label{s9a}
\begin{array}{c}
\displaystyle{
e^{-D_{\alpha}(a)D_{\alpha}(b)F+(\p_{\alpha}-\p_{\beta})^2F}
\Bigl (\frac{a}{b}
e^{-(D_{\alpha}(a)-D_{\alpha}(b))(\p_{\alpha}-\p_{\beta})F} -
\frac{b}{a}
e^{(D_{\alpha}(a)-D_{\alpha}(b))(\p_{\alpha}-\p_{\beta})F}\Bigr )}
\\ \\
=(a-b)(D_{\alpha}(a)+D_{\alpha}(b))\p_{t_{\beta ,1}}F.
\end{array}
\eeq
The limit of equation (\ref{s10}):
\beq\label{s10a}
\epsilon_{\alpha \beta}e^{(\p_{\alpha}-\p_{\mu}+D_{\alpha}(a))
(\p_{\beta}-\p_{\mu}+D_{\beta}(b))F}=
\epsilon_{\alpha \mu}\epsilon_{\beta \mu}\Bigl (\p_{\beta}+D_{\beta}(b)-
\p_{\alpha}-D_{\alpha}(a)\Bigr )\p_{t_{\mu, 1}}F.
\eeq
As is proved in \cite{TT07}, these equations are equivalent to the
universal Whithem hierarchy for genus zero.

Let us rewrite these equations in a more suggestive form. 
For this purpose, we introduce the notation
\beq\label{d4}
R_{\alpha}=e^{\p_{\alpha}^2F}, \qquad 
R_{\alpha \beta}=R_{\beta \alpha}=e^{\p_{\alpha}\p_{\beta}F}
\eeq
and the functions
\beq\label{d5}
\begin{array}{l}
\displaystyle{
w_{\alpha}(z)=ze^{-D_{\alpha} (z)\p_{\alpha}F -
\p_{\alpha}^2F},}
\\ \\
\displaystyle{
w_{\alpha \beta}(z)=e^{-D_{\alpha} (z)\p_{\beta}F
-\p_{\alpha}\p_{\beta}F}},
\end{array}
\eeq
\beq\label{d6}
\begin{array}{l}
p_{\alpha}(z)=z -(\p_{\alpha}+D_{\alpha}(z))
\p_{t_{\alpha , 1}}F,
\\ \\
p_{\alpha \beta}(z)=-(\p_{\alpha}+D_{\alpha}(z))
\p_{t_{\beta , 1}}F
\end{array}
\eeq
(note that $R_{\alpha \alpha}=R_{\alpha}$ but 
$w_{\alpha \alpha}(z)\neq w_{\alpha}(z)$, 
$p_{\alpha \alpha}(z)\neq p_{\alpha}(z)$).
In this notation, the equations (\ref{s4a})--(\ref{s9a}) read as follows:
\beq\label{s4b}
e^{D_{\alpha}(a)D_{\alpha}(b)F}=\frac{p_{\alpha}(a)-p_{\alpha}(b)}{a-b},
\eeq

\beq\label{s5b}
e^{D_{\alpha}(a)D_{\alpha}(b)F}=\frac{R_{\alpha}}{R_{\alpha \beta}}\,
\frac{w_{\alpha}(a)w_{\alpha \beta}^{-1}(a)-
w_{\alpha}(b)w_{\alpha \beta}^{-1}(b)}{a-b},
\eeq

\beq\label{s6b}
e^{D_{\alpha}(a)D_{\beta}(b)F}=1+\frac{w_{\alpha \beta}(a)
w_{\beta \alpha}(b)}{w_{\alpha}(a)w_{\beta}(b)},
\eeq

\beq\label{s7b}
\frac{R_{\alpha}}{R_{\alpha \beta}}
\, \frac{w_{\alpha}(a)}{w_{\alpha \beta}(a)}\,
e^{D_{\alpha}(a)D_{\beta}(b)F}=p_{\alpha}(a)-p_{\beta \alpha}(b),
\eeq

\beq\label{s8b}
R_{\mu \alpha}^{-1}w_{\mu \alpha}^{-1}(a)-
R_{\mu \beta}^{-1}w_{\mu \beta}^{-1}(a)+
\epsilon_{\alpha \beta}\epsilon_{\alpha \mu}\epsilon_{\beta \mu}
R_{\alpha \beta}R_{\mu \alpha}^{-1}R_{\mu \beta}^{-1}
w_{\mu}^{-1}(a)=0,
\eeq

\beq\label{s9b}
\begin{array}{l}
\displaystyle{
\frac{R_{\alpha}R_{\beta}}{R^2_{\alpha \beta}}\left (
\frac{w_{\alpha} (a)w_{\alpha \beta}(b)}{w_{\alpha} (b)w_{\alpha \beta} (a)}
-\frac{w_{\alpha}(b)w_{\alpha \beta}(a)}{w_{\alpha} (a)w_{\alpha \beta} (b)}
\right )e^{-D_{\alpha}(a)D_{\alpha}(b)F}}
\\ \\ \phantom{aaaaaaaaaaaaaaaaaaa}
=-(a-b)(p_{\alpha \beta}(a)+
p_{\alpha \beta}(b)-2p_{\alpha \beta}(\infty )),
\end{array}
\eeq

\beq\label{s10b}
\epsilon_{\alpha \beta}\frac{R_{\mu}}{R_{\alpha \beta}}
\, \frac{w_{\alpha \mu}(a)w_{\beta \mu}(b)}{w_{\alpha \beta}(a)
w_{\beta \alpha}(b)}\, e^{D_{\alpha}(a)D_{\beta}(b)F}=
\epsilon_{\alpha \mu}\epsilon_{\beta \mu}
(p_{\alpha \mu}(a)-p_{\beta \mu}(b)).
\eeq
In the next section we show that these seven equations can be reduced
to one.

\section{The dispersionless equations in trigonometric form}

Dividing equations (\ref{s4b}), (\ref{s5b}), we get the relation
$$
p_{\alpha}(a)-\frac{R_{\alpha}w_{\alpha}(a)}{R_{\alpha \beta}
w_{\alpha \beta}(a)}=
p_{\alpha}(b)-\frac{R_{\alpha}w_{\alpha}(b)}{R_{\alpha \beta}
w_{\alpha \beta}(b)},
$$
from which it follows that 
$\displaystyle{p_{\alpha}(z)-\frac{R_{\alpha}w_{\alpha}(z)}{R_{\alpha \beta}
w_{\alpha \beta}(z)}}$ does not depend on $z$. The limit $z\to \infty$
yields:
\beq\label{t1}
p_{\alpha}(z)-\frac{R_{\alpha}w_{\alpha}(z)}{R_{\alpha \beta}
w_{\alpha \beta}(z)}=p_{\beta \alpha}(\infty ).
\eeq

Next, multiply equations (\ref{s4b}) and (\ref{s9b}) and express
$w_{\alpha}$ through $p_{\alpha}$ 
in the left hand side with the help of (\ref{t1}). This yields
the relation
$$
R_{\alpha \beta}^2 (p_{\alpha \beta}(a)-p_{\alpha \beta}(\infty ))+
\frac{R_{\alpha}R_{\beta}}{p_{\alpha}(a)-p_{\beta \alpha}(\infty )}
=-R_{\alpha \beta}^2 (p_{\alpha \beta}(b)-p_{\alpha \beta}(\infty ))-
\frac{R_{\alpha}R_{\beta}}{p_{\alpha}(b)-p_{\beta \alpha}(\infty )},
$$
from which it follows that
\beq\label{t2}
R_{\alpha \beta}^2 p_{\alpha \beta}(z)+
\frac{R_{\alpha}R_{\beta}}{p_{\alpha}(z)-p_{\beta \alpha}(\infty )}=
R_{\alpha \beta}^2 p_{\alpha \beta}(\infty ).
\eeq
This equation, connecting $p_{\alpha \beta}$ and $p_{\alpha}$, defines a rational curve. 

It is natural to uniformize this curve using trigonometric
functions. For this purpose, we introduce a function $u_{\alpha}(z)$
normalized so that $u_{\alpha}(\infty )=0$ with expansion around 
$\infty$ of the form
\beq\label{t3}
u_{\alpha}(z)=\frac{c_{\alpha ,1}}{z}+\sum_{k\geq 2}
\frac{c_{\alpha ,k}}{z^k},
\eeq
with the coefficients depending on the times. 
In terms of this function the uniformization reads:
\beq\label{t4}
\begin{array}{l}
\displaystyle{
p_{\alpha}(z)=\gamma_{\alpha}\frac{\cos u_{\alpha}(z)}{\sin u_{\alpha}(z)},}
\\ \\
\displaystyle{
p_{\alpha \beta}(z)=\gamma_{\beta}
\frac{\cos (u_{\alpha}(z)+\eta_{\alpha \beta})}{\sin 
(u_{\alpha} (z)+\eta_{\alpha \beta})},}
\\ \\
R_{\alpha}=\gamma_{\alpha}, \quad R_{\alpha \beta}=
\sin \eta_{\alpha \beta}.
\end{array}
\eeq
Here $\gamma_{\alpha}=\gamma_{\alpha}({\bf t})$,
$\eta_{\alpha \beta}=\eta_{\alpha \beta}({\bf t})$ are some functions
of the times. Note that it should be $R_{\alpha \beta}=R_{\beta \alpha}$
but, as we shall see later, the assumption that $\eta_{\alpha \beta}=
\eta_{\beta \alpha}$ is wrong. Instead,
\beq\label{t5}
\eta_{\beta \alpha}=\pi - \eta_{\alpha \beta}.
\eeq
With this relation, substitution of (\ref{t4}) 
into (\ref{t2}) brings the latter to identity.

Plugging (\ref{t4}) into (\ref{t1}), one finds:
\beq\label{t6}
\begin{array}{l}
\displaystyle{
w_{\alpha}(z)=\frac{1}{\sin u_{\alpha}(z)},}
\\ \\
\displaystyle{
w_{\alpha \beta}(z)=\frac{1}{\sin (u_{\alpha} (z)+\eta_{\alpha \beta})}}.
\end{array}
\eeq
Tending $z\to \infty$ in the first of equations (\ref{t4}), 
we get the relation
\beq\label{t7}
\gamma_{\alpha}({\bf t})=c_{\alpha , 1}({\bf t}).
\eeq

In the trigonometric parametrization equations (\ref{s4b}), (\ref{s5b})
and
(\ref{s6b}), (\ref{s7b}), (\ref{s10b})
become the same. In order to write them in a compact form, we introduce
the differential operator
\beq\label{t8}
\nabla_{\alpha}(z)=\p_{\alpha}+D_{\alpha}(z).
\eeq
The equations read:
\beq\label{t9}
e^{\nabla_{\alpha}(a)\nabla_{\alpha}(b)F}=
\frac{\sin (u_{\alpha}(a)-u_{\alpha}(b))}{a^{-1}-b^{-1}},
\eeq
\beq\label{t10}
e^{\nabla_{\alpha}(a)\nabla_{\beta}(b)F}=
\sin (u_{\alpha}(a)-u_{\beta}(b)+\eta_{\alpha \beta}).
\eeq
(In the second equation it is assumed that $\alpha \neq \beta$.)
In the limit $b\to \infty$ they become
\beq\label{t11}
ae^{\nabla_{\alpha}(a)\p_{\alpha}F}=
\sin u_{\alpha}(a),
\eeq
\beq\label{t12}
e^{\nabla_{\alpha}(a)\p_{\beta}F}=
\sin (u_{\alpha}(a)+\eta_{\alpha \beta}).
\eeq

Finally, we consider equation (\ref{s8b}). In the trigonometric
parametrization the dependence on $a$ disappears and the
equation becomes a constraint for $\eta_{\alpha \beta}$:
\beq\label{t13}
\epsilon_{\alpha \beta}\sin \eta_{\alpha \beta}=
\epsilon_{\alpha \mu}\epsilon_{\beta \mu}\sin (\eta_{\mu \alpha} -
\eta_{\mu \beta}),
\eeq
which should hold for all distinct $\alpha , \beta , \mu$. 
The solution is
\beq\label{t14}
\eta_{\alpha \beta}=\eta_{\alpha} -\eta_{\beta}+
\frac{\pi}{2}(\epsilon_{\alpha \beta}+1)
\eeq
with some $\eta_{\alpha}$.

The final result can be summarized as follows. Let us redefine
the functions $u_{\alpha}(z)$ including into them 
the constant terms in the expansion as $z\to \infty$. So we introduce
the functions $v_{\alpha}(z)=u_{\alpha}(z)+\eta_{\alpha}$. 
Then all equations of the hierarchy are encoded in the single 
equation
\beq\label{t15}
\epsilon_{\beta \alpha}e^{\nabla_{\alpha}(a)\nabla_{\beta}(b)F}=
\frac{\sin (v_{\alpha}(a)-v_{\beta}(b))}{
(a^{-1}-b^{-1})^{\delta_{\alpha \beta}}}
\eeq
valid for all complex $a,b$. 

The meaning of this equation is that general second order derivatives
of the function $F$ in the independent variables, as is usual in the 
dispersionless equations, are expressed through
some special second order derivatives. Indeed, the equation for $\alpha
\neq \beta$ can be written as
\beq\label{t16}
\epsilon_{\beta \alpha}e^{\nabla_{\alpha}(a)\nabla_{\beta}(b)F}
=\sin \Bigl ( \eta_{\alpha}-\eta_{\beta}+\mbox{arcsin}\, (a^{-1}
e^{\nabla_{\alpha}(a)\p_{\alpha}F})-
\mbox{arcsin}\, (b^{-1}
e^{\nabla_{\beta}(b)\p_{\beta}F})\Bigr ).
\eeq
At the same time, since
$$
\begin{array}{ll}
\displaystyle{
\eta_{\alpha}-\eta_{\beta}=\sum_{\gamma =\alpha}^{\beta -1} (\eta_{\gamma}
-\eta_{\gamma +1}),} & \quad \alpha <\beta ,
\\ & \\
\displaystyle{
\eta_{\alpha}-\eta_{\beta}=-\sum_{\gamma =\beta}^{\alpha -1} (\eta_{\gamma}
-\eta_{\gamma +1}),} & \quad \alpha >\beta ,
\end{array}
$$
we have:
$$
\begin{array}{ll}
\displaystyle{
\eta_{\alpha}-\eta_{\beta}=-\sum_{\gamma =\alpha}^{\beta -1} 
\mbox{arcsin} (e^{\p_{\gamma}\p_{\gamma +1}F}),}
& \quad \alpha <\beta ,
\\ & \\
\displaystyle{
\eta_{\alpha}-\eta_{\beta}=
\sum_{\gamma =\beta}^{\alpha -1} 
\mbox{arcsin} (e^{\p_{\gamma}\p_{\gamma +1}F}),}
& \quad \alpha >\beta .
\end{array}
$$
Therefore, as equation (\ref{t16}) shows, the general 
second order derivatives of the function $F$ are expressed through the
special derivatives $\p_{\alpha}^2 F$, $\p_{\alpha}\p_{\alpha +1}F$,
$\p_{t_{\alpha ,k}}\p_{\alpha}F$. 

Finally, by applying $\nabla_{\gamma}(c)$ to the logarithm of both
sides of (\ref{t15}), this equation can be written as
\beq\label{t17}
\begin{array}{c}
\nabla_{\alpha}(a)\log \sin (v_{\beta}(b)-v_{\gamma}(c))=
\nabla_{\beta}(b)\log \sin (v_{\gamma}(c)-v_{\alpha}(a))
\\ \\
=\nabla_{\gamma}(c)\log \sin (v_{\alpha}(a)-v_{\beta}(b)).
\end{array}
\eeq
This symmetry is a manifistation of integrability of the 
dispersionless multicomponent KP
hierarchy.

\section{Concluding remarks}

We have reconsidered the dispersionless limit of the multicomponent 
KP hierarchy. Nonlinear differential equations for the dispersionless
limit of logarithm of the tau-function have been obtained. 
We have shown that there is a rational curve built in the structure
of the hierarchy. This curve can be uniformized via trigonometric
functions. 
In the 
trigonometric parametrization, the equations of the hierarchy acquire
especially nice form, being encoded in a single equation. 

This work was motivated by our study of the multicomponent DKP 
hierarchy (work in progress). In the fermionic approach, the tau-function
of the KP hierarchy 
is an expectation value of exponent of neutral quadratic form in fermions
while for the DKP hierarchy (known also as the Pfaff lattice) 
the quadratic form can be arbitrary. In this
sense the latter is more general than the former. In the dispersionless
limit of the DKP hierarchy there is an elliptic curve built 
in its structure \cite{Takasaki07,Takasaki09}, 
with the elliptic modulus being a dynamical variable. Uniformizing
this curve via elliptic functions, one can represent the equations
in a nice elliptic form (for the one-component case see \cite{AZ14}). 
Basically, the $\sin$-function in (\ref{t15}) and other 
equations is replaced 
by the elliptic $\mbox{sn}$-function. We have found it instructive
to reconsider the dispersionless KP hierarchy within a similar approach,
where elliptic functions degenerate to trigonometric ones.

\section*{Acknowledgments}

\addcontentsline{toc}{section}{Acknowledgments}

The author thanks K.Takasaki and T.Takebe for reading the manuscript
and valuable comments. 
This work is an output of a research project 
implemented as a part of the Basic Research Program 
at the National Research University Higher School 
of Economics (HSE University).

\end{document}